# The statistics of the ordering of chiral ribbons on a honeycomb lattice

Max M. McCarthy[1], William S. Fall[1,], Xiangbing Zeng[2], Goran Ungar[2,3] and Gillian A Gehring[1]

1. Department of Physics and Astronomy, University of Sheffield, Sheffield S3 7RH, UK.
2. Department of Materials Science and Engineering, University of Sheffield, Sheffield S1 3JD, UK.
3. State Key Laboratory for Mechanical Behavior of Materials, Shaanxi International Research Centre for Soft Materials, Xi'an Jiaotong University, Xi'an, 710049, China.

## Abstract

A novel model, devised to describe spontaneous chirality synchronization in complex liquids and liquid crystals, is proposed and studied. Segments of ribbon-like molecular columns with left- or right-handed 180° twist lie on the bonds of a honeycomb lattice so that three ribbons meet in a vertex of the hexagonal honeycomb. The energy of each vertex is a minimum if the three ribbons have the same chirality, $-\varepsilon$, and is $+\varepsilon$ otherwise, and the ground state is homochiral, i.e. all ribbons have the same handedness. The energy levels for two vertices linked by a single ribbon are either $-2\varepsilon$, 0 and $+2\varepsilon$ in this vertex model. Monte Carlo simulations demonstrate that this model is identical to an Ising spin model on a Kagome lattice, for which the site energy structure is quite different. The equivalence of the ordering of the vertex and Ising spin models is also shown analytically. The energy difference between the disordered and ground states, $4J$ in the spin model, is related to the transition temperature for the Kagome lattice using the exact result, $T_c=2.14J$. The ordering energy difference for a single site is 50% higher for the vertex model. The thermodynamic energy for the vertex model is corrected by a factor of 1/3 due to double counting and this makes the specific heat of the vertex model also equal to that of the spin model as expected. Other similar models where there is an unusual relation between the site and thermodynamic energies are discussed briefly.

**I Introduction**

Chirality is ubiquitous in nature as it is built into the molecules that sustain life, such as proteins, DNA and carbohydrates, as well as in most modern drugs. For example the alpha-helix that form proteins is right handed because natural amino acids are left handed. However, even molecules without intrinsic chirality can form uniform helices. Most achiral but crystallisable (isotactic, syndiotactic) synthetic polymers crystallize in a helical conformation, with long-range helical order sustained by close packing with neighbouring chains.[1,2] Furthermore, most instantaneous conformations of achiral molecules are chiral but time-averaging in a fluid phase renders them achiral. If the energy barrier between enantiomeric conformations is relatively high, a small amount of a chiral dopant can tip the balance and makes one handedness prevail. This is known as chirality amplification, or the sergeants and soldiers effect.[3] It is promoted by helix formation either through self-assembly in columns in solution forming a gel [4,5] or in non-crystalline but helix-forming polymers with high conformational barriers such as poly(phenyl acetylenes).[6,7] In both cases the barriers for chiral interconversion of the individual molecule is compounded by close packing in a helical column, making the switch in chirality a cooperative process.[8,9] If no chiral impurity or chiral surface are present, such chirogenic molecules can tip either way, sometimes resulting in multidomain samples of random handedness,[10] often referred to as a "dark conglomerate"[11].

A few years ago, studies of thermotropic (i.e. solvent-free) liquid crystals have revealed that one of the two most common bicontinuous cubic phases, the triple-network phase previously thought to have spacegroup $Im\bar{3}m$,[12,13] is always chiral and optically active, even if it contains only achiral molecules.[14] Recent re-examination reassigned it to a lower symmetry *I*23, still retaining the triple network cubic nature but with a modified structure (figure 1b,e).[15] Conversely the other common cubic, the double gyroid $Ia\bar{3}d$ phase (figure 1a), is never chiral. Another LC phase, known as Smectic-Q, has since also been found to belong to the same family of bicontinuous phases but is tetragonal and also shows spontaneous chirality in achiral compounds.[16] The typical molecules forming these phases consist of a rod-like aromatic core with between 1 and 3 flexible chains attached at the ends, typically alkyl or oligo(ethylene oxide). These rod-like molecules lie normal to the segment axis in rafts of 3 or 4 (figure 1c). To explain the chirality of such bicontinuous phases it was proposed that the network segments are ribbon-like and chiral due to the twist in molecular orientation, in all cases by 7-9°, in successive rafts along the segment (Figure 1d). This twist angle is a compromise between the tendency of the aromatic cores of neighbouring molecules to stay parallel and maximize their π-π interaction, while avoiding the steric clash between their bulky molten end-chains. To allow for smooth convergence and close packing, it is proposed that all three or four ribbon-like network segments maintain the same twist sense at a network junction, resulting in macroscopic homochirality. Incidentally, the reason that the double gyroid phase shows no outward chirality is that its two networks have opposite twist sense, cancelling each other's chirality[11]. Such cancellation is not possible in the triple network $I23$ phase. Furthermore, in the Smectic-Q phase the two networks are of the same hand, hence the phase is always optically active.

Even more surprising than the above finding of spontaneous chirality synchronization in bicontinuous LC phases was the discovery that in some of these compounds macroscopic chirality and strong optical activity are maintained even in the isotropic liquid, which we refer to as Iso* phase, above the isotropization temperature $T_i$ of the LC.[17] Iso* then transforms to the ordinary optically inactive liquid (Iso), typically up to 20 K above $T_i$ through a well defined transition at $T_c$ that appears to be second order.[18] The structure of Iso* is uncertain, but since its optical rotation at $T_i$ is similar to that of the $I23$ cubic, it is likely that it also contains networks with twisted segments and 3-way vertices, only without the long-range positional order. An overview of a significant number of compounds exhibiting the double gyroid and the triple network phases, all based on twisted ribbons, is given in [19]. The chiral isotropic phase has recently been reported in a number of quite diverse compounds with rod-like shape.[20,21,22]

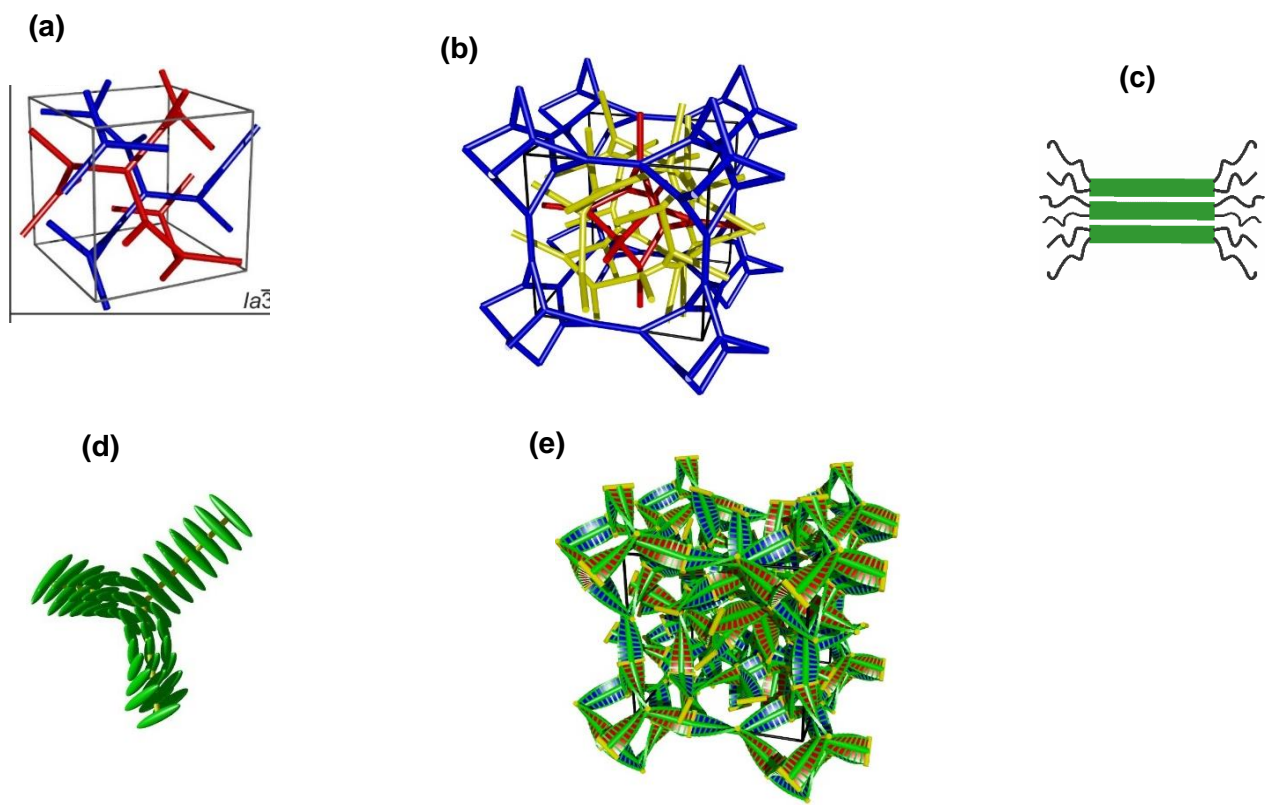

**Figure 1.** Models of a unit cell of the **(a)** double network gyroid cubic phase $Ia\bar{3}d$ and **(b)** the triple network cubic *I*23. **(c)** A raft of 3 rod-like molecules; green: aromatic core, black: flexible chains. **(d)** A three-way junction (or vertex) showing how all three segments must have the same twist sense in order for the rod-like molecular cores (green) from the three segments to merge smoothly. **(e)** The ribbon representation of the *I*23 phase where each green rib represents a raft **(c)** of 3-4 molecules lying perpendicular to the ribbon axis. a,b,d,e reproduced from 15 by permission of the Royal Society of Chemistry.

We model the Iso\*-Iso transition by considering the statistical mechanics of a system with three-way planar vertices constrained to lie on a two-dimensional honeycomb lattice. In this model we assume that all ribbons connecting two adjoining vertices are twisted by ±180° with chirality ±1 and neglect the possibility of energetically costly helix reversals within the ribbon segment.

The vertex energy is favourable, $-\varepsilon$, if the three ribbons meeting at a vertex have the same chirality and is unfavourable, with energy $+\varepsilon$, if the chirality of any one ribbon differs from the other two. This condition is the driving force that can cause a given ribbon to switch chirality. This is an interesting statistical system because there are clearly more unfavourable states of a single vertex, 6, compared with 2 favourable states. This contrasts with the connecting ribbons that have only two states, chirality ±1. A phase transition occurs from a disordered state to a state where one chirality dominates and the material becomes uniformly optically active.

We show that the statistical properties of the vertex model on a honeycomb lattice map exactly on to an Ising model on a kagome lattice with the same transition temperature with one dramatic difference, namely that the average site energy for the vertex model is 50% larger than given for the spin model at high temperatures. We also consider other models where the transition temperature is unrelated to the site energy of ordering as is found here.

**II Energetics of the vertex model compared with the spin model**

The ribbons lie on honeycomb lattice vectors as shown in figure 2(a) with vertices at the lattice points. The chirality of each lattice vector is represented by a pseudo-spin $\sigma = \pm 1$ as shown by a red dot in figure 2(a). The blue dots are the three–way vertices. If lines are drawn between the red dots in figure 2(a) they form a kagome lattice - the dual of the honeycomb lattice. The magnetic properties of ferromagnetically coupled spins on a kagome lattice are exactly solvable by an extension to Onsager's solution, with critical temperature $T_c = 2.1433 J$ [23]. The pseudo-spins of the vertex model in two dimensions lie on the kagome lattice but the situation differs from the ferromagnetic spin model because the energies of the vertices involve three spin interactions.

The site energy corresponding to the five-spin cluster shown in Figure 2(b) takes one of three values: $-2\varepsilon$ if both vertices are in the favourable state, $+2\varepsilon$ if both vertices are in the unfavourable state and 0 if the state of one vertex is favourable and the other unfavourable. In Table 1 the energies of the site 0 in the configuration $(\sigma_1 \sigma_2 \boldsymbol{\sigma_0} \sigma_3 \sigma_4)$ are compared with those for a spin model with ferromagnetic exchange $J$ and spins ±1 shown in figure 2(c). The multiplicities are given by a factor of 2 by reversing all spins and permuting the spins 1-4. In each case there are $2^5$=32 states. We find the average site energy which is the energy the system will reach in the limit of high temperatures is $+\varepsilon$.

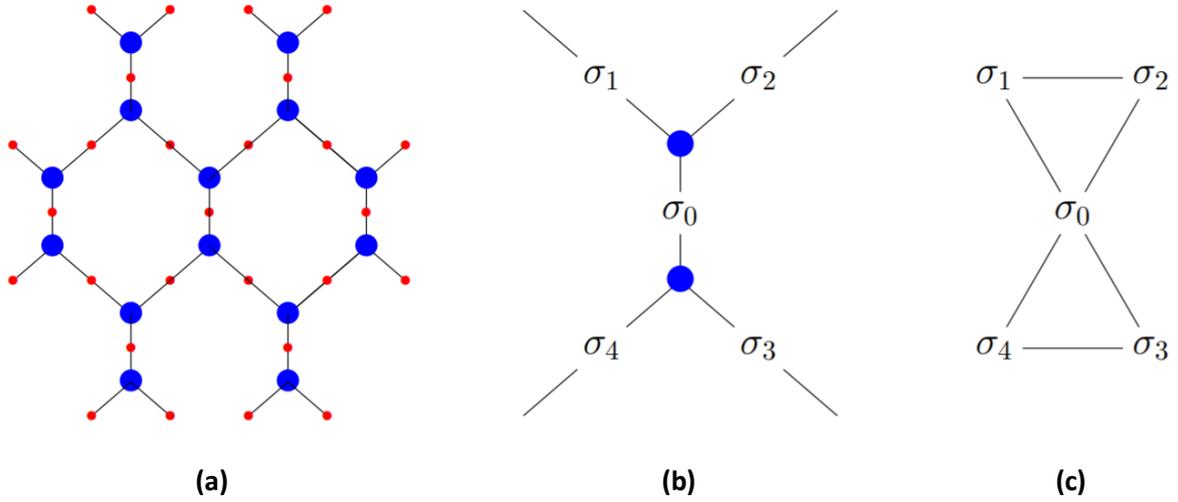

|       (a)       |       (b)       |       (c)       |

**Figure 2**. (**a**) shows a honeycomb lattice with the vertices as blue dots and the chiral ribbons marked with red dots; (**b**) shows a five pseudo-spin unit centered on spin **0** where the spin-spin interaction occurs though the energies of the two vertices; (**c**) a kagome spin model where the energy of spin at site 0 is given by $-J\sigma_0(\sigma_1+\sigma_2+\sigma_3+\sigma_4)$.

Table I

The energies of the five site cluster for the vertex model and the ferromagnetic spin model on the kagome lattice

| Configuration $\sigma_1\sigma_2\boldsymbol{\sigma_0}\sigma_3\sigma_4$ | Energy Vertex Model | Energy Spin Model | Multiplicity |
|---|---|---|---|
| + + **+** + + | $-2\varepsilon$ | $-4J$ | 2 |
| + + **-** + + | $+2\varepsilon$ | $+4J$ | 2 |
| + + **+** - - | 0 | 0 | 4 |
| +- **+** +- | $+2\varepsilon$ | 0 | 8 |
| +- **+** + + | 0 | $-2J$ | 8 |
| +- **-** + + | $+2\varepsilon$ | $+2J$ | 8 |
| Total | $+32\varepsilon$ | 0 | 32 |
| Average energy | $+\varepsilon$ | 0 | |

In spin models the difference between the average energy, always zero, and the ground state energy divided by the difference between the highest and lowest energy is equal to one half. The value for the vertex model is ¾. This may be compared with an *n*-state Potts model where the value is $\frac{n-1}{n}$ so the vertex model might have some similarity to the 4-state Potts which is known to have a second order transition[24,25].

The distribution of energies and their multiplicities of the central site are shown in Figure 3.

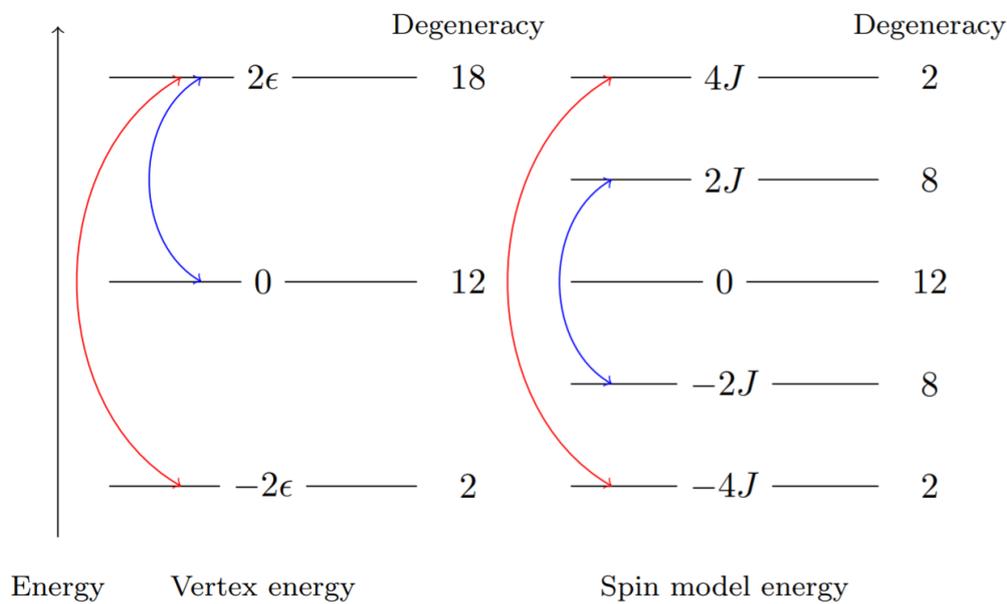

**Figure 3**. The energies and multiplicity of the central site for (a) the vertex and (b) the spin model. The red and blue lines link the states where transitions occur if the central spin is reversed.

### III Monte Carlo simulation of the vertex model

The simulations are run using a Metropolis algorithm and Glauber dynamics over honeycomb lattices of varying size using the energetics given in Table I. For comparison the calculations were repeated for the spin model. At each temperature the routine was run $10^4$ times to establish equilibrium and each measurement was the average over $10^6$ iterations. All the runs were done as the sample was heated from low temperature. Energy changes for different configurations, *E*(config.1) - $E$(config.2), are shown in Table II for both models. The transitions are shown in Figure 3 as red for the larger and blue for the smaller energy.

If the energy change is zero the spin is flipped in the Monte Carlo code with probability ½ . This was chosen rather than the standard procedure of accepting the move when the energy change was zero because it improved the convergence of the average site energy at high temperatures for the vertex model. It is shown in Table II that a move of zero energy can occur for an initial vertex state of energy $2\varepsilon$ or zero; whereas zero energy moves occur for the spin model only when the initial energy state is zero. We believe that this difference was responsible for the slow convergence at high temperatures that we observed for the vertex model and not the spin model.

**Table II** The change in energy after the central spin is reversed in vertex and spin models, $\Delta E_V$ and $\Delta E_S$

| Configuration1 $\sigma_1\sigma_2\ \sigma_0\ \sigma_3\sigma_4$ $E_V$    $E_S$ | Configuration2 $\sigma_1\sigma_2\ \sigma_0\ \sigma_3\sigma_4$ $E_V$    $E_S$ | $\Delta E_V$ | $\Delta E_S$ |
|---|---|---|---|
| + +  **+**  + +  <br> $-2\varepsilon$   $-4J$ | + +  -  + +  <br> $+2\varepsilon$   $+4J$ | $-4\varepsilon$ | $-8J$ |
| + +  **+**  - -  <br> 0        0 | + +  -  - -  <br> 0        0 | 0 | 0 |
| +-  **+**  + -  <br> $+2\varepsilon$     0 | +-  -  + -  <br> $+2\varepsilon$     0 | 0 | 0 |
| +-  **+**  + +  <br> 0        $-2J$ | +-  -  + +  <br> $+2\varepsilon$   $+2J$ | $-2\varepsilon$ | $-4J$ |

The results in Table II show that the energy of configuration 1 minus that for configuration 2 are *identical* for the same change in orientation of the central spin for vertex and the spin models provided we choose $2J = \varepsilon$. This means that the results for the magnetisation, $m(T) = <\sigma>$, the susceptibility, and Binder cumulant will be identical for these two models.

The susceptibility is given by,

$$\chi(T) = \frac{1}{k_B T} \sum_{i,j} \left[ \langle \sigma_i \sigma_j \rangle - \langle \sigma_i \rangle \langle \sigma_j \rangle \right]. \qquad [1]$$

For systems that have a second order transition the maximum value of the susceptibility, $\chi_{max}$, and lattice size $L$ (sizes of 20, 40, 80, 160 and 320 were used in the Monte-Carlo simulations) are related by $\chi_{max} \sim L^y$ where $y=\gamma/\nu=7/4$ for the spin model[26]. Near the transition the value of susceptibility is highly sensitive to small temperature changes hence the temperature steps were reduced to $0.005\,\varepsilon$ in the critical region.

Plots of the temperature dependence of the susceptibility, $\chi$, are shown in figure 4(a) for the vertex model. In this and in all subsequent figures $T$ is given in units of $J$. The spin and vertex models have identical susceptibilities, hence the purpose of these plots is to validate our Monte Carlo procedures and Figure 4(b) shows a log-log plot of $\chi_{max}$ as a function of $L$. It is seen that the Monte Carlo simulations for the vertex model agree extremely well with the known behaviour of the Ising model. The exponent $p$ in the expression $\chi_{max} \sim L^p$ is found from the plot in figure 4(b) to be $p$=1.766 which is a little higher than the Ising exponent, $\gamma/\nu = 7/4$.

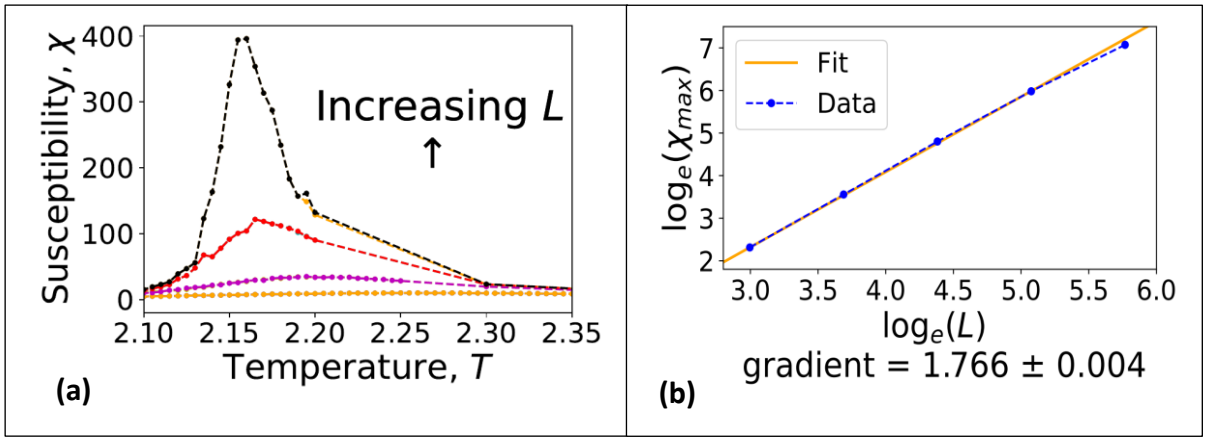

**Figure 4.** Susceptibility plots for the vertex model to confirm its identity to the spin model. **(a)** temperature dependence of the susceptibility as a function of temperature for different lattice sizes $L$ =20, 40, 80, 160 **(b)** plot of $\log_e \chi_{max}$ as a function of $\log_e L$, the slope for the spin model is 7/4 =1.75

The most accurate way to determine the transition temperature is to use a Binder cumulant[27] that is defined for a lattice of size *L* by,

$$U_L = 1 - \frac{<\sigma^4>}{3<\sigma^2>^2} \quad .$$

[2]

$U_L$ is independent of *L* at a second order transition. The approximation for the transition temperature is found from the intersections of $U_L$ as a function of temperature, for different *L*. The transition temperature for the vertex model is identical to that of the spin model, 2.1433[17], within the error as illustrated in Figure 5 bounding it above and below by 2.146 and 2.143 with an error of 0.003 respectively.

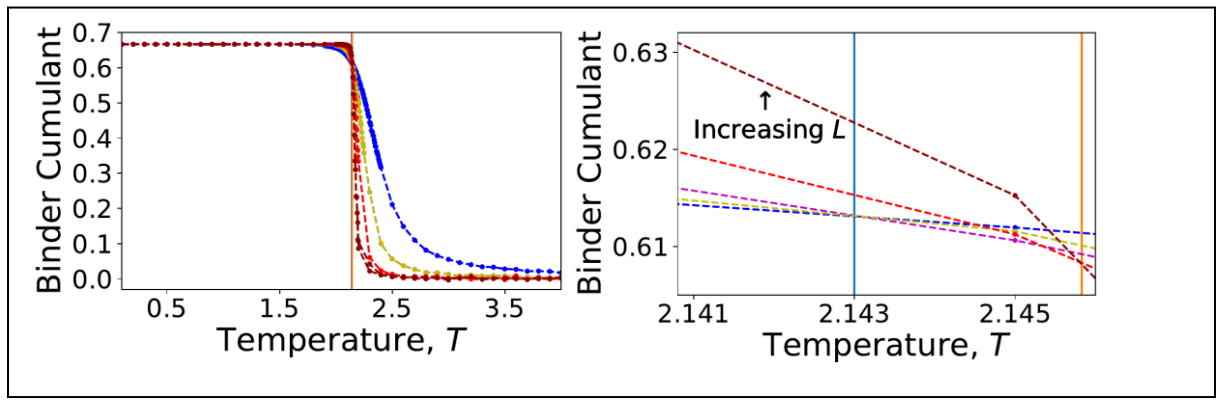

**Figure 5.**: The Binder cumulants for the vertex model with *L* =20, 40, 80,160 **(a)** over the full temperature range and **(b)**, over a restricted temperature range near to $T_c$. The vertical lines indicate the range of possible values of $T_c$.

So far we have exact agreement bewween the vertex and the spin models but the energy values show a markedly different story. The ground state energy, $-2\varepsilon$, is the same in both cases if $\varepsilon = 2J$. However the energies in the high temperature limit are very different being zero for the spin model but $+\varepsilon$ for the vertex model. The difference in energy between the ground and fully disordered states is $3\varepsilon$ for the vertex model compared to $4J = 2\varepsilon$ for the spin model. Values for the site energy as a function of temperature for the vertex and spin models are shown in figure 6(a) where it is clear the high temperature limits are different. This apparent discrepancy is discussed in more detail in section V.

## V A unified formulation for the vertex and spin model

The site energy for the vertex model may be rewritten in terms of pseudo-spins, $\sigma = \pm 1$ using the relations that $(\sigma_0 + \sigma_1 + \sigma_2)^2 = 9/1$ for identical/dissimilar spins.

$$E(\text{site0}) = -A\left[(\sigma_0 + \sigma_1 + \sigma_2)^2 + (\sigma_0 + \sigma_3 + \sigma_4)^2 + C\right] \quad [4]$$

The values of the parameters, A and C, are found by requiring that the maximum and minimum energies are given by $\pm 2\varepsilon$, hence $A = \dfrac{\varepsilon}{4}$ and $C = -10$.

This gives equation [5] which demonstrates the relationship between the vertex and the spin model.

$$\begin{aligned}E(\text{site0}) &= -\frac{\varepsilon}{4}\left[(\sigma_0 + \sigma_1 + \sigma_2)^2 + (\sigma_0 + \sigma_3 + \sigma_4)^2 - 10\right] \\ &= -\frac{\varepsilon}{2}\left[\sigma_0(\sigma_1 + \sigma_2 + \sigma_3 + \sigma_4) + (\sigma_1\sigma_2 + \sigma_3\sigma_4) - 2\right]\end{aligned} \quad [5]$$

The site energy for the vertex model, $E_{site}(0)$, clearly differs from that of the spin model by the addition of the terms $\sigma_1\sigma_2 + \sigma_3\sigma_4$. These do not contribute to the change in energy when the spin $\sigma_0$ is flipped and hence are irrelevant in a Monte Carlo calculation. However they do contribute to the values of $<E_{site}(0)>$ as a function of temperature as shown below in figure 6a.

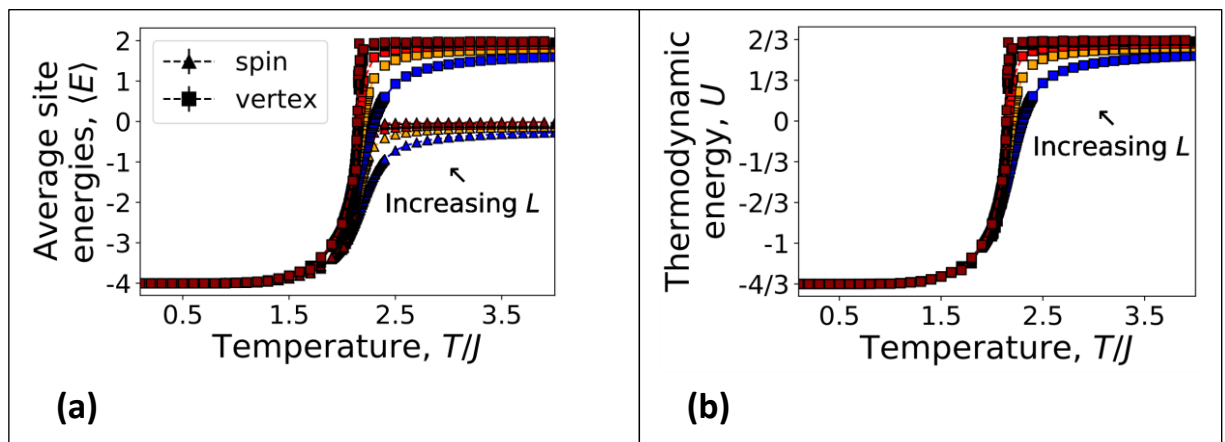

(a)           (b)

**Figure 6. a)** A plot of the average site energies $<E>$ of both models with a triangle representing the spin model and a square the vertex for $L$ =20, 40, 80,160, 360 **(b)** The thermodynamic energy of the vertex model calculated for $L$ =20, 40, 80,160, 360.

The thermodynamic energy, $U$, for the spin model is given by $U = \frac{1}{2N}\sum_n E_{site}(n)$ where the factor ½ is present to prevent the double counting of the correlation functions e.g. $<\sigma_n \sigma_{n+\rho}>$. A different correction for multiple counting occurs in the vertex model. The correlation functions $<\sigma_0\sigma_1 + \sigma_1\sigma_2 + \sigma_2\sigma_0>$ will be present in the three site energies, $E_{site}(0)$, $E_{site}(1)$ and $E_{site}(2)$, as can be seen from figure 2(b). This will be true for all vertices hence the thermodynamic energy is given by $U = \frac{1}{3N}\sum_n E_{site}(n)$ for the vertex model. This gives the energy shown in figure 6(b).

The specific heats of the vertex model are identical to those of the spin model as shown in figure 7.

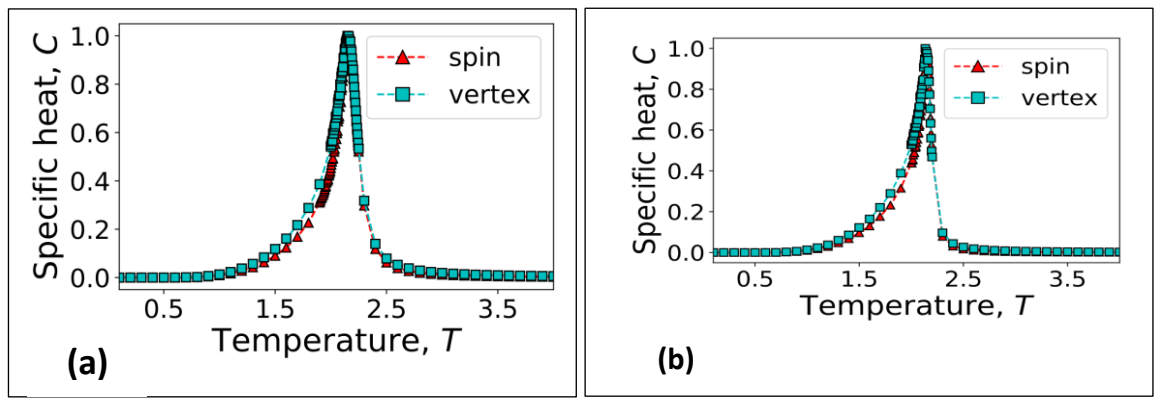

(a)  (b)

**Figure. 7**: Plots of the specific heat as a function of temperature, an increased $L$ gives a reduced divergence in $C$. **(a)**, **(b)**, geometric similarity for both the vertex and spin models, for $L$=40 and 80.

A continuum of models may be defined for $0 \leq \alpha \leq 1$,

$$E_\alpha(\text{site0}) = -\frac{\varepsilon}{2}\left[\sigma_0(\sigma_1 + \sigma_2 + \sigma_3 + \sigma_4) + \alpha(\sigma_1\sigma_2 + \sigma_3\sigma_4 - 2)\right] \qquad [7]$$

The energies and degeneracies of this model are shown in figure 8. The maximum and minimum energies are independent of α; however the energy of the fully disordered state is at $+\varepsilon$ for the vertex model and at zero for the spin model.

The thermodynamic energy for this model is given by,

$$U_\alpha = \frac{1}{(2+\alpha)N} \sum_n E_\alpha(\text{site } n).$$

All the thermodynamic properties of these models are independent of $\alpha$.

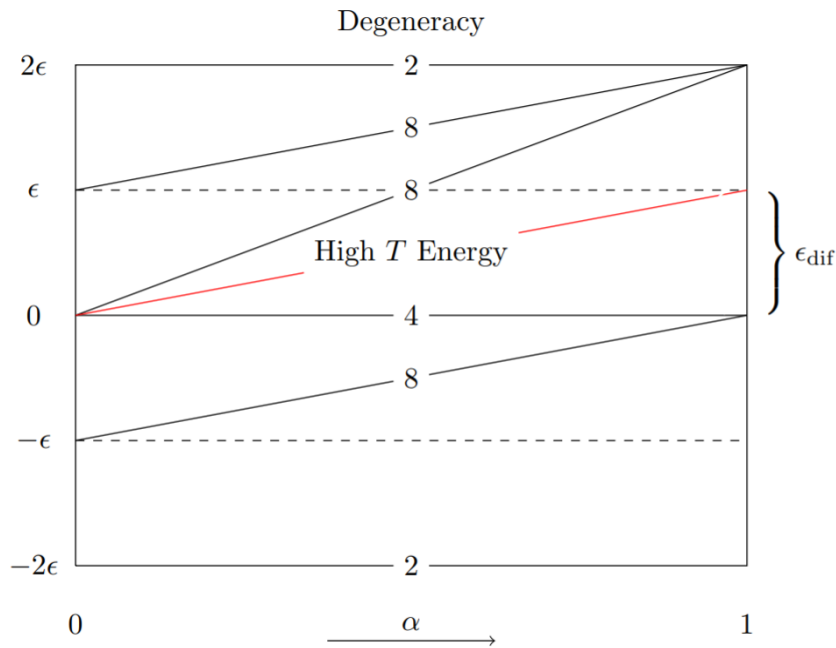

**Figure 8**. The site energies as a function of $\alpha$. The energies for the spin model are at $\alpha = 0$ and for the vertex model at $\alpha = 1$.

**VI Other similar models with an unusual relation between the site energy and the thermodynamic energy**

The site energy for the vertex model given in equation [4] contained terms that did not depend on the central site. These terms contributed to the ordering energy but not to the value of $T_c$. We consider what other models would show this feature.

A four-way vertex model would map on to a triangular lattice as shown in figure 9. The vertex energy would have three values corresponding to the states where all 4 ribbons have the same chirality, the state where the chirality of one ribbon is different and the state where there are equal numbers of each chirality.

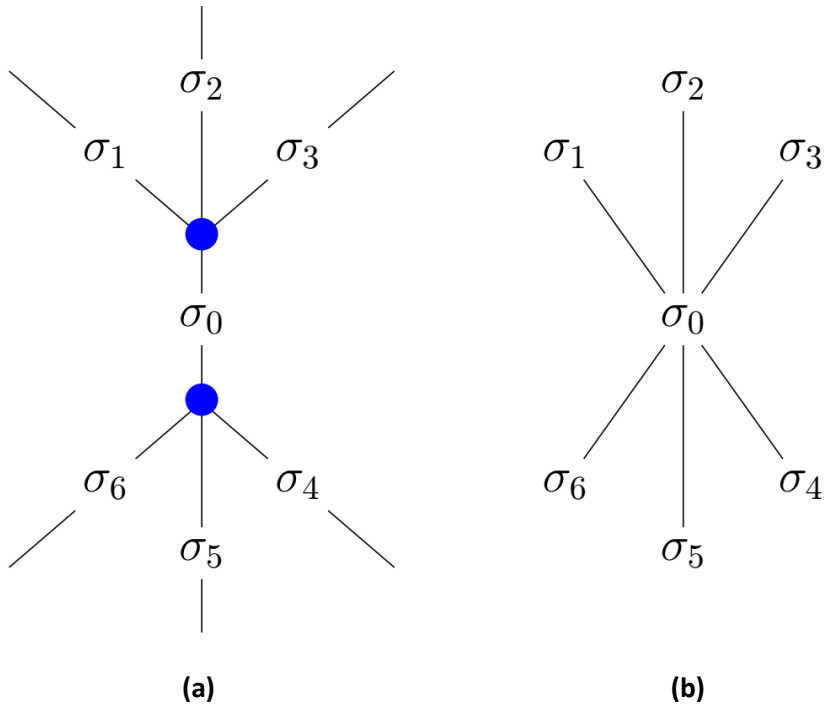

**Figure 9** The vertex model for a 4-way vertex, **(a),** and the corresponding triangular lattice model, **(b)**.

We chose the vertex energy in this case to be $-\gamma(\sigma_1+\sigma_2+\sigma_3+\sigma_4)^2$ which gives the three energies of the vertex to be $-16\gamma, -4\gamma$ and 0. This would give the ordering energy to be as follows:

$$E(site0) = -\gamma\left[(\sigma_0+\sigma_1+\sigma_2+\sigma_3)^2 + (\sigma_0+\sigma_4+\sigma_5+\sigma_6)^2\right]$$
$$= -2\gamma\left[\sigma_0(\sigma_1+\sigma_2+\sigma_3+\sigma_4+\sigma_5+\sigma_6) + (\sigma_1\sigma_2+\sigma_2\sigma_3+\sigma_3\sigma_1+\sigma_4\sigma_5+\sigma_5\sigma_6+\sigma_6\sigma_4) + 8\right]$$

[8]

The contributions to the ordering energy at $T=0$ are $12\gamma$ from the terms involving spin $\sigma_0$ and an equal contribution from the additional terms. In this case the elimination of double counting will require that the relationship between the thermodynamic energy and the site energies is given by $U_{4\text{vertex}} = \frac{1}{4N} \sum_n E_\alpha(\text{site } n)$.

Another way of introducing extra terms is to consider multiple spin interactions[28] as in the site energy below for a square lattice:

$$E(\text{site}0) = -J\left[\sigma_0(\sigma_1 + \sigma_2 + \sigma_3 + \sigma_4) + \delta\sigma_1\sigma_2\sigma_3\sigma_4\right]. \tag{9}$$

The ordering energy for this model is $J(4+\delta)$ but Monte Carlo simulations will predict that the transition is independent of δ. The negative fourth order term would tend to drive the transition to first order; this occurs for $\delta > 1/3$ within mean field theory applied to this two dimensional lattice.

The models described here have simple solutions as they are not frustrated as occurs when antiferromagnetic interactions are involved. Such frustrated models have been discussed extensively and include the ice-model and the 8-vertex model which may also map on to solvable models[29].

**VII Conclusion**

A vertex model was developed to describe the ordering of spontaneously chiral liquid Iso[*] and network-based bicontinuous liquid crystals in non-chiral compounds where the ordering energy arises from chiral ribbons linking in three-way vertices. The centres of the ribbons lie on a kagome lattice and the thermodynamic properties of the vertex model were identical to that of the two dimensional spin model even though the allowed energies of a single site were radically different from those of the spin model on a kagome lattice.

The theory developed here allows us to estimate the energy difference between vertices in Iso* for which all three connecting ribbons are and are not identical, $2\varepsilon$. Using the Iso* – Iso transition at 465K and R$T_c$ =2.1433$J$ for the Kagome lattice *we* find that $J$=0.430 kcal/mol and since $\varepsilon$=2$J$ we find $\varepsilon$=0.86 kcal/mol. This is likely to be an overestimate for $\varepsilon$ because it was obtained from the combination of a phase transition temperature for a three dimensional system with a two dimensional model.

The observed transition is indeed second order[18] and the energy of the transition may be estimated from an integral over the specific heat peak. This experimental result gives an estimate of the ordering energy, $U = 1.1$ kcal/mol which is higher than the value of $\varepsilon = 0.86$ kcal/mol that would be deduced using this model. We conclude that the two dimensional model of the three dimensional Iso* – Iso transition is qualitatively correct although the detailed numbers are subject to error. However this vertex model does have interesting statistical properties in its own right.

**References**

[1] Lotz, B; Wittmann, JC; Lovinger, (1996) AJ; Structure and morphology of poly(propylenes): A molecular analysis *Polymer* **37**, 4979-4992.

[2] De Rosa, C and. Auriemma, F. Editors *Crystals and Crystallinity in Polymers: Diffraction Analysis of Ordered and Disordered Crystals* John Wiley & Sons 2013.

[3] Roche, C; Sun, H.-J.; Prendergast, M.; Leowanawat, P.; Partridge, B.; Heiney, P. A.;. Araoka, F.; Graf, R.; Spiess, H. W.; Zeng, X.B.; Ungar, G.; Percec V.; Homochiral Columns by Chiral Self-Sorting During Supramolecular Helical Organisation of Hat-Shaped Molecules (2014) *J. Am. Chem. Soc.* **136**, 7169–7185

[4] Shen, Z.C; Jiang, Y.Q.; Wang, T.Y.; Liu, M.H. (2015) Symmetry Breaking in the Supramolecular Gels of an Achiral Gelator Exclusively Driven by π−π Stacking *J. Am. Chem. Soc.* **137**, 16109−16115

[5] Jonkheijm, P;. van der Schoot, P; Schenning, A. P. H. J.;. Meijer,( 2006) E. W Probing the Solvent-Assisted Nucleation Pathway in Chemical Self-Assembly. *Science* **313**, 80-83.

[6] Yashima, E.; Maeda K. and Okamoto Y (1999) Memory of macromolecular helicity assisted by interaction with achiral small molecules *Nature* **39**, 449.

[7] Yashima,E.; Ousaka,N.; Taura,D.; Shimomura,K.; Ikai,T. and Maeda K. (2016) Supramolecular Helical Systems: Helical Assemblies of Small Molecules, Foldamers, and Polymers with Chiral Amplification and Their Functions *Chem. Rev.*, **116**, 13752−13990

[8] Malthete J. and Collet A. (1987) Inversion of the Cyclotribenzylene Cone in a Columnar Mesophase: A Potential Way to Ferroelectric Materials. *J. Am. Chem. Soc.* **109,** 7545

[9] Green M.M., Park J. W.,. Sato T.,. Teramoto A., Lifson S,. Selinger R. L. B, Selinger J. V. (1999) The macromolecular route to chiral amplification, *Angew. Chem. Int. Ed.* **38**, 3138-3154.

[10] Tschierske C. and Ungar G. (2016) "Mirror-Symmetry Breaking by Chirality Synchronization in Liquids and Liquid Crystals of Achiral Molecules", *Chem Phys Chem*, **17**, 9-26

[11] Takezoe H., (2011)*Top. Curr. Chem.*, **318**, 303–330.

[12] Levelut A. M. & Clerc, M. (1998) Structural investigations on 'smectic D' and related mesophases. *Liquid Crystals* **24,** 105–115

[13] Zeng X.B., Ungar G., Impéror-Clerc M (2005) A Triple-Network Tricontinuous Cubic Liquid Crystal, *Nature Materials*, 4 562-567.

[14] Dressel, C.; Liu, F.; Prehm, M.; Zeng, X.; Ungar, G.; Tschierske, C. (2014). Dynamic Mirror-Symmetry Breaking in Bicontinuous Cubic Phases. *Angew. Chem. Int. Ed.* **53**, pp.13115-13120.

[15]. Zeng X. B,. Ungar G (2020)Spontaneously Chiral Cubic Liquid Crystal: Three Interpenetrating Networks with a Twist , *J. Mater. Chem. C* **8**, 5389-5398

[16] Lu H. J.,. Zeng X. B, Ungar G, Dressel C,. Tschierske C (2018) Solution of the Puzzle of Smectic-Q: The Phase Structure and the Origin of Spontaneous Chirality, *Angew. Chem. Int. Ed.* **57**, 2835-2840

[17] Dressel, C.; Reppe, T.; Prehm, M.; Brautzsch, M.; Tchierske, C. (2014). `Chiral self-sorting and amplification in isotropic liquids of achiral molecules'. Nature Chemistry,. **6**, pp.971-977.

[18] Huanjun L, (2018) PhD Thesis, University of Sheffield.

[19] Dressel C, Reppe T., Poppe S.,. Prehm M,. Lu H.J., Zeng X.B., Ungar, G. Tschierske C. (2020) Helical networks of π-conjugated rods - A robust design concept for bicontinuous cubic liquid crystalline phases with achiral Ia3d and chiral I23 lattice *Adv. Funct. Mater.* **30**, 2004353.

[20] Dressel C., Weissflog W., Tschierske C. (2015) Spontaneous mirror symmetry breaking in a re-entrant isotropic liquid. *Chem. Commun.,* **51**, 15850.

[21] Alaasar M., Prehm M., Cao Y., Liu F., Tschierske C. (2016) Spontaneous Mirror-Symmetry Breaking in Isotropic Liquid Phases of Photoisomerizable Achiral Molecules. *Angew. Chem. Int. Ed.,* **55**, 312 –316.

[22] Alaasar M., Poppe S., Dong Q., Liu F., Tschierske C. (2017) Isothermal Chirality Switching in Liquid-Crystalline Azobenzene Compounds with Non-Polarized Light. *Angew. Chem. Int. Ed.* **56**, 10801 –10805.

[23] Fisher, M.E. (1958). Transformations of Ising Models. *Phys Rev.***113,** pp.969-981

[24] M P M den Nijs (1979) A relation between the temperature exponents of the eight-vertex and q-state Potts model *J. Phys. A: Math. Gen.* **12** 1857

[25] Wu F.Y. (1982) The Potts model *Rev. Mod. Phys*, **54**, pp235-268

[26] Crokidakis N (2009 ) First-order phase transition in a 2D random-field Ising model with conflicting dynamics *J. Stat. Mech.* **2009**, 02058.

[27] Binder K. (1981) Finite size scaling analysis of Ising model block distribution functions *Z. Phys. B Cond. Matt.* **43**, 119-140

[28] Johnston D .A., Mueller M., Janke W. (2017) Plaquette Ising models, degeneracy and scaling *Eur. Phys. J. Special Topics* **226**, 749–764

[29] Baxter R.J. Exactly solved models in Statistical Mechanics Academic Press 1982 pages 202- 321